\documentclass[aps,showpacs,amsmath,amssymb,12pt]{revtex4}
\usepackage{latexsym}
\usepackage{bm}

\begin{document}

\title{G{\"o}del-Type Metrics in Various Dimensions}
\author{Metin G{\"u}rses}
\email{gurses@fen.bilkent.edu.tr}
\affiliation{Department of Mathematics, Faculty of Sciences,\\
             Bilkent University, 06800, Ankara, Turkey}

\author{Atalay Karasu}
\email{karasu@metu.edu.tr}
\affiliation{Department of Physics, Faculty of Arts and  Sciences,\\
             Middle East Technical University, 06531, Ankara, Turkey}

\author{{\"O}zg{\"u}r Sar{\i}o\u{g}lu}
\email{sarioglu@metu.edu.tr}
\affiliation{Department of Physics, Faculty of Arts and  Sciences,\\
             Middle East Technical University, 06531, Ankara, Turkey}

\date{\today}

\begin{abstract}
G{\"o}del-type metrics are introduced and used in producing
charged dust solutions in various dimensions. The key ingredient
is a $(D-1)$-dimensional Riemannian geometry which is then
employed in constructing solutions to the Einstein-Maxwell field
equations with a dust distribution in $D$ dimensions. The only
essential field equation in the procedure turns out to be the
source-free Maxwell's equation in the relevant background.
Similarly the geodesics of this type of metric are described by
the Lorentz force equation for a charged particle in the lower
dimensional geometry. It is explicitly shown with several examples
that G{\"o}del-type metrics can be used in obtaining exact
solutions to various supergravity theories and in constructing
spacetimes that contain both closed timelike and closed null
curves and that contain neither of these. Among the solutions that
can be established using non-flat backgrounds, such as the
Tangherlini metrics in $(D-1)$-dimensions, there exists a class
which can be interpreted as describing black-hole-type objects 
in a G{\"o}del-like universe.
\end{abstract}


\maketitle

\section{\label{intro} Introduction}

G{\"o}del's metric \cite{god} in general relativity is the solution of
Einstein's field equations with homogeneous perfect fluid distribution
having $G_{5}$ maximal symmetry \cite{kram}. This spacetime admits
closed timelike and closed null curves but contains no closed timelike
and null geodesics \cite{eh}. The G{\"o}del universe is geodesically
complete, and doesn't contain any singularities or horizons. There
have been several attempts to generalize the G{\"o}del metric in
classical general relativity \cite{reti,kant,rogo,bada}. The main goal
of these works has been the elimination of closed timelike and closed
null curves.

We call a metric in $D$ dimensions as a {\it G{\"o}del-type metric} 
if it can be written in the form $g_{\mu\nu}=h_{\mu\nu}-u_{\mu}u_{\nu}$ 
where $u^{\mu}$ is a timelike unit vector and $h_{\mu\nu}$ is a 
degenerate matrix of rank $D-1$ with the additional condition that 
$h_{\mu \nu}$ be the metric of an Einstein space of a 
$(D-1)$-dimensional Riemannian geometry.

In fact, taken at face value, such a decomposition of spacetime metrics  
has of course been adopted by several researchers with various aims.
These are generally called 3+1 decompositions in General Relativity.
One well known work is due to Geroch \cite{gero} in $D=4$ where our
$u^{\mu}$ is taken as \( u^{\mu} = \xi^{\mu} / \sqrt{|\lambda|} \) in 
which $\xi^{\mu}$ is a Killing vector field to start with and 
$\lambda = \xi_{\mu} \xi^{\mu}$. However, Geroch does not put 
any restrictions on the 3-dimensional metric $h_{\mu\nu}$ unlike our
case. Geroch reduces the vacuum Einstein field equations to a scalar,
complex, Ernst-type nonlinear differential equation and develops a 
technique for generating new solutions of the vacuum Einstein field
equations from vacuum spacetimes. Although the G{\"o}del-type metrics 
we define and use here are of the same type, it must be kept in mind
that our $h_{\mu \nu}$ is the metric of an Einstein space of a 
$(D-1)$-dimensional Riemannian geometry. We also do not assume $u^{\mu}$
to be a Killing vector field to start with, but with the other restrictions
we impose it turns out to be one. Another major difference is that 
we look for all possible $D$-dimensional G{\"o}del-type metrics,
and hence $u^{\mu}$ vectors, that produce physically acceptable matter
content for Einstein field equations.

Metrics of this form also look like the well known Kerr-Schild metrics
of classical general relativity \cite{kerr} which have
$g_{\mu\nu}=\eta_{\mu\nu} - \ell_{\mu} \ell_{\nu}$ for a null
vector $\ell^{\mu}$ and which we have recently used in
constructing accelerated Kerr-Schild geometries for the Einstein-Maxwell 
null dust \cite{gs1}, Einstein-Born-Infeld null dust field
equations \cite{gs2}, and their extensions with a cosmological
constant and respective zero acceleration limits \cite{gs3}.

Remarkably the very form of the G{\"o}del-type metrics is also
reminiscent of the metrics used in Kaluza-Klein reductions in
string theories \cite{duff}. However, as will be apparent in the
subsequent sections, G{\"o}del-type metrics have a number of
characteristics that distinguish them from the Kaluza-Klein metrics.
Here the background metric $h_{\mu\nu}$ is taken as positive definite
whereas in the Kaluza-Klein case it must be locally Lorentzian.
Moreover, contrary to what is done in the Kaluza-Klein mechanism,
the G{\"o}del-type metrics are used in obtaining a $D$-dimensional
theory starting from a $(D-1)$-dimensional one. The $D$-dimensional
timelike vector $u^{\mu}$ is used in the construction of a Maxwell
theory in $D$ dimensions unlike the Kaluza-Klein vector potential
which lives and defines a Maxwell theory in $D-1$ dimensions. Even
though G{\"o}del-type metrics are akin to metrics employed in
the Kaluza-Klein mechanism at face value, the applications we present
here should make their real worth clear and should help in contrasting
them with Kaluza-Klein metrics.

G{\"o}del-type metrics also show up in supergravity theories in
some dimensions. A special class of G{\"o}del-type metrics is
known to be the T-dual of the pp-wave metrics in string theory
\cite{calk,bghv,ht}. These metrics are all supersymmetric but contain
closed timelike and closed null curves and thus violate causality
\cite{gh,gghp,her1,her2,hik}. Recently there has been an
attempt to remedy this problem by introducing observer dependent
holographic screens \cite{bghv,bdebr}. In \cite{dfsim}, a new
class of supergravity solutions has been constructed which locally look
like the G{\"o}del universe inside a domain wall made out of supertubes
and which do not contain any closed timelike curves. There have also been
studies that describe black holes embedded in G{\"o}del spacetimes
\cite{gh,her1,greg} and brane-world generalizations of the G{\"o}del
universe \cite{bats}.

In this work, we consider G{\"o}del-type metrics in a
$D$-dimensional spacetime manifold $M$. We show that in all
dimensions the Einstein equations are classically equivalent to
the field equations of general relativity with a charged dust
source provided that a simple $(D-1)$-dimensional Euclidean source-free 
Maxwell's equation is satisfied. The energy density of the
dust fluid is proportional to the Maxwell invariant $F^{2}$. We
next show that the geodesics of the G{\"o}del-type metrics are
described by solutions of the $(D-1)$-dimensional Euclidean
Lorentz force equation for a charged particle. We then discuss the 
possible existence of examples of spacetimes containing closed timelike 
and closed null curves which violate causality and examples of spacetimes 
without any closed timelike or closed null curves where causality is
preserved. We show that the G{\"o}del-type metrics we introduce
provide exact solutions to various kinds of supergravity theories
in five, six, eight, ten and eleven dimensions. All these exact 
solutions are based on the vector field $u_{\mu}$ which satisfies 
the $(D-1)$-dimensional Maxwell's equation in the background of some 
$(D-1)$-dimensional Riemannian geometry with metric $h_{\mu\nu}$. In 
this respect, we do not give only a specific solution but in fact 
provide a whole class of exact solutions to each of the aforementioned 
theories. We construct some explicit examples when $h_{\mu \nu}$ is 
trivially flat, i.e. the identity matrix of $D-1$ dimensions.

We next consider an interesting class of the G{\"o}del-type
metrics by taking a $(D-1)$-dimensional non-flat background
$h_{\mu\nu}$. We specifically consider the cases when the background 
$h_{\mu\nu}$ is conformally flat, an Einstein space and, as a subclass, 
a Riemannian Tangherlini solution. We explicitly construct such examples 
for $D=4$ even though these can be generalized to dimensions $D>4$ as 
well. When the background is an Einstein space, the corresponding source 
for the Einstein equations in $D$ dimensions turns out to be a charged
perfect fluid with pressure density $p=\frac{1}{2}(3-D) \Lambda$
(so that $p>0$ when $\Lambda<0$) and energy density
$\rho=\frac{1}{4} f^2 + \frac{1}{2} (D-1) \Lambda$, where
$\Lambda$ is the cosmological constant and $f^2$ denotes the
Maxwell invariant. We also discuss the existence of closed
timelike and closed null curves in this class of spacetimes and
explicitly construct geometries with and without such curves in
$D=4$. We show that when the background is a Riemannian
Tangherlini space, the $D$-dimensional solution turns out to
describe a black-hole-type object depending on the parameters.
We then finish off with our conclusions and a discussion of
possible future work.

\section{\label{asilgodmet} G\"{o}del-Type Metrics}

Let $M$ be a $D$-dimensional manifold with a metric of the form
\begin{equation}
g_{\mu \nu}= h_{\mu \nu}-u_{\mu} u_{\nu} \; . \label{met}
\end{equation}
Here $h_{\mu \nu}$ is a degenerate $D \times D$ matrix with rank equal to
$D-1$. We assume that the degeneracy of $h_{\mu\nu}$ is caused by 
taking $h_{k\mu}=0$, where $x^{k}$ is a fixed coordinate with 
$0 \le k \le D-1$ (note that $x^{k}$ does not necessarily have to be 
spatial), and by keeping the rest of $h_{\mu\nu}$, i.e. 
$\mu \neq k$ or $\nu \neq k$, dependent on all the coordinates 
$x^{\alpha}$ except $x^{k}$ so that $\partial_{k} h_{\mu\nu}=0$.
Hence, in the most general case, ``the background'' $h_{\mu\nu}$ can 
effectively be thought of as the metric of a $(D-1)$-dimensional 
non-flat spacetime. As for $u^{\mu}$, we assume that it is 
a timelike unit vector, $u_{\mu} u^{\mu}=-1$, and that $u_{\mu}$ is 
independent of the fixed special coordinate $x^{k}$, i.e.
$\partial_{k} u_{\mu}=0$. These imply that one can take
$u^{\mu}=-\frac{1}{u_{k}}\, \delta^{\mu}_{\;k}$. 

Now the question we ask is as follows: Let us start with a metric of the
form (\ref{met}) and calculate its Einstein tensor. Can the Einstein
tensor be interpreted as describing the energy momentum tensor of a 
physically acceptable source? Does one need further assumptions on
$h_{\mu\nu}$ and/or $u_{\mu}$ so that ``the left hand side'' of 
$G_{\mu\nu} \sim T_{\mu\nu}$ can be thought of giving an acceptable
``right hand side'', i.e. corresponding to a physically
reasonable matter source? As you will see in the subsequent sections,
the answer is ``yes'' provided that one further demands $h_{\mu \nu}$ 
to be the metric of an Einstein space of a $(D-1)$-dimensional 
Riemannian geometry. We call such a metric $g_{\mu\nu}$ a 
{\it G\"{o}del-type of metric}. The sole reason we use this name is
because of the fact that some of the spacetimes we find also have 
closed timelike curves and some of the supergravity solutions we 
present have already appeared in the literature with a title referring
to G\"{o}del.

In the most general case, $u_{k} \neq constant$ and the assumptions 
we have made so far show that $u^{\mu}$ is not a Killing vector. However 
if one further takes $u_{k} = constant$, then it turns out to be one. 
Throughout this work we will assume that $u_{k} = constant$. We will
first consider the simple case of $h_{\mu\nu}$ being flat. For this case,
we will examine what can be said and done in classical general relativity
in the remaining parts of this section and investigate how one can use
flat backgrounds to find solutions to various supergravity theories 
in Section \ref{vardims}. We will consider the case of non-flat backgrounds 
later in Section \ref{nonflat}.

\subsection{\label{godmet} Solutions of Einstein's Equations 
in Flat Backgrounds}

Throughout the rest of this section and in Section \ref{vardims}, we 
will further assume that $h_{0\mu}=0$, $h_{ij}=\bar{\delta}_{ij}$, 
the $(D-1)$-dimensional Kronecker delta symbol and
$\partial_{\alpha} h_{\mu\nu}=0$. We take Greek indices to run from $0,1,
\cdots$ to $D-1$ whereas Latin indices range from $1$ to $D-1$. 
(Our conventions are similar to the conventions of Hawking and Ellis 
\cite{eh}.) The determinant of $g_{\mu\nu}$ is then $g=-u_{0}^2$ and moreover
$u^{\mu}=-\frac{1}{u_{0}}\, \delta^{\mu}_{0}$. In what follows, we will also
assume that $u_{0}=1$ and that $\partial_{0}\, u_{\alpha}=0$.

With these assumptions, it is not hard to show that $u^{\mu}\, h_{\mu \nu}=0$
and the inverse of the metric can be calculated to be
\begin{equation}
g^{\mu \nu}=\bar{h}^{\mu\nu}+(-1+\bar{h}^{\alpha\beta} \, u_{\alpha}\,
u_{\beta}) \, u^{\mu} \, u^{\nu}
+ u^{\mu} (\bar{h}^{\nu\alpha} \, u_{\alpha}\,)
+ u^{\nu} (\bar{h}^{\mu\alpha} \, u_{\alpha}\,) \; . \label{invmet}
\end{equation}
Here $\bar{h}^{\mu\nu}$ is the $(D-1)$-dimensional inverse of $h_{\mu \nu}$;
i.e. \( \bar{h}^{\mu \nu} \, h_{\nu \alpha} = \bar{\delta}^{\mu}_{\;\alpha}\)
with $\bar{\delta}^{\mu}_{\;\alpha}$ denoting the $(D-1)$-dimensional
Kronecker delta: \( \delta^{\mu}_{\;\alpha} = \bar{\delta}^{\mu}_{\;\alpha} +
\delta^{\mu}_{\;0} \, \delta^{0}_{\;\alpha} \).

The Christoffel symbols can now be calculated to be
\begin{equation}
\Gamma^{\mu}\,_{\alpha \beta}=\frac{1}{2}\, (u_{\alpha} \, f^{\mu}\,_{\beta}
+ u_{\beta} \, f^{\mu}\,_{\alpha}) - \frac{1}{2} \, u^{\mu} \,
(u_{\alpha, \, \beta} + u_{\beta, \, \alpha}) \label{chris}
\end{equation}
where we have used $f_{\alpha \beta} \equiv u_{\beta, \, \alpha} -
u_{\alpha, \, \beta}$; $u_{\alpha, \, \beta} \equiv \partial_{\beta}
\, u_{\alpha}$ and $f^{\mu}\,_{\nu}=g^{\mu \alpha}\, f_{\alpha \nu}$.
We will also use a semicolon to denote a covariant derivative with respect
to the Christoffel symbols given above; $u_{\alpha ; \, \beta} \equiv
\nabla_{\beta} \, u_{\alpha}$. One can easily show that
\( u^{\alpha} \, u_{\beta; \, \alpha} = 0 \) and 
\(  u_{\beta; \, \alpha} = \frac{1}{2} f_{\alpha \beta}, \)
hence $u^{\mu}$ is tangent to a timelike geodesic curve and is a timelike
Killing vector.

The Ricci tensor can be calculated to be
\begin{equation}
R_{\mu \nu}=\frac{1}{2}\, f_{\mu}\,^{\alpha}\, f_{\nu \alpha}
-\frac{1}{2}\,(u_{\mu}\, j_{\nu}+u_{\nu}\, j_{\mu})
+\frac{1}{4}\,f^2 \, u_{\mu} \, u_{\nu} \, \label{ricci}
\end{equation}
where we have used $f^2 \equiv f^{\alpha \beta}\,f_{\alpha \beta}$ and
$j_{\mu} \equiv \partial_{\alpha}\, f_{\mu}\,^{\alpha}$. [Note that it is
not possible to have $j_{\mu}=k u_{\mu}$ for a nontrivial constant $k$.
Suppose the contrary is true, i.e. $j_{\mu}=k u_{\mu}$. Now by 
\( u^{\alpha} \, f_{\mu \alpha} = 0 \), one finds
\( \partial_{\alpha} (u^{\mu} \, f_{\mu}\,^{\alpha}) = 0 .\)
However $u^{\mu} = - \delta^{\mu}_{0}$ and this gives $u^{\mu} j_{\mu}=0$
which implies that $j_{0}=k=0$.] The Ricci scalar is then easily found to be
\begin{equation}
R = \frac{1}{4}\, f^2 - u^{\mu} \, j_{\mu} \; . \label{riccis}
\end{equation}
Choosing $j_{\mu}=0$, the Einstein tensor is simply given by
\begin{equation}
G_{\mu \nu}=\frac{1}{2}\, T^{f}_{\mu \nu} + \frac{1}{4}\, f^{2}\,
u_{\mu} \, u_{\nu} \, ,
\label{eint}
\end{equation}
where
\( T^{f}_{\mu \nu} \equiv f_{\mu \alpha} \, f_{\nu}\,^{\alpha} - \frac{1}{4} \,
g_{\mu\nu} \, f^{2} \)
is the Maxwell energy-momentum tensor for
$f_{\mu \nu}$. Obviously, (\ref{eint}) implies that the G{\"o}del
type-metric $g_{\mu\nu}$ (\ref{met}) is a solution of the charged
dust field equations in $D$ dimensions. The energy density of the dust
fluid is just $\frac{1}{4} f^{2}$ then. Now one only needs to make sure
that $j_{\mu}=0$ is satisfied. However, this implies that
\begin{equation}
\partial_{i}\, f_{i j}=0 \, , \label{max}
\end{equation}
with our choice of $h_{\mu\nu}$ and $u_{\mu}$. Hence all that is left to
solve is the flat $(D-1)$-dimensional Euclidean source-free Maxwell's equation.
In covariant form (\ref{max}) can also be written as
\begin{equation}
J^{\nu} \equiv \nabla_{\mu} \, f^{\mu\nu} = \frac{1}{2}\, f^{2}\, u^{\nu} \; .
\label{cmax}
\end{equation}
Indeed (\ref{cmax}) will be useful for the remaining part of this work.

A few remarks regarding the positivity of energy and the character of the
geodesics are in order at this point. For a timelike vector $\xi^{\mu}$,
one has $T^{f}_{\mu \nu} \, \xi^{\mu} \, \xi^{\nu} \ge 0$ by the very nature
of $T^{f}_{\mu \nu}$ and since $f_{0\mu}=0$, one has $f^2=(f_{ij})^2 \ge 0$
as well. Hence it is readily seen that
\[ G_{\mu \nu} \, \xi^{\mu} \, \xi^{\nu} =
T^{f}_{\mu \nu} \, \xi^{\mu} \, \xi^{\nu}
+\frac{1}{4} \, f^2 \, (u_{\mu} \xi^{\mu})^{2} \ge 0 \, , \]
for all timelike $\xi^{\mu}$ and the weak energy condition is satisfied
for spacetimes described by G{\"o}del-type metrics.

As for the behavior of the geodesics, let us start by taking a geodesic
curve $\Gamma$ on $M$ which is parametrized as $x^{\mu}(\tau)$. Using
(\ref{chris}) and denoting the derivative with respect to the affine parameter
$\tau$ by a dot, the geodesic equation yields
\[ \ddot{x}^{\mu} + f^{\mu}\,_{\beta} \, \dot{x}^{\beta} \,
(u_{\alpha} \, \dot{x}^{\alpha}) - u^{\mu} \, \dot{x}^{\alpha}
(u_{\alpha, \, \beta} \, \dot{x}^{\beta}) = 0 \, . \]

Noting that $u_{\alpha, \, \beta} \, \dot{x}^{\beta} = \dot{u}_{\alpha}$,
writing $f^{\mu}\,_{\beta}$ explicitly via the inverse of the metric
(\ref{invmet}) and using \( u^{\alpha} f_{\mu\alpha} = 0 \), 
this becomes
\begin{equation}
\ddot{x}^{\mu} + u_{\alpha} \, \dot{x}^{\alpha} \,
(\bar{h}^{\mu\sigma} + u^{\mu} \, \bar{h}^{\sigma\nu} \, u_{\nu}) \,
f_{\sigma\beta} \, \dot{x}^{\beta} - u^{\mu} \,
(\dot{u}_{\alpha} \, \dot{x}^{\alpha}) = 0 \, , \label{geod1}
\end{equation}
and contracting this with $u_{\mu}$, one obtains
a constant of motion for the geodesic equation as
\begin{equation}
u_{\mu} \, \dot{x}^{\mu} = \dot{x}^{0} + u_{i} \, \dot{x}^{i}
= -e = constant \, . \label{geodch}
\end{equation}
Meanwhile setting the free index $\mu=i$ in (\ref{geod1}), one also finds
\( \ddot{x}^{i} - e \, (\bar{h}^{i\sigma} \, f_{\sigma\beta} \,
\dot{x}^{\beta}) = 0 \, , \)
or simply
\begin{equation}
\ddot{x}^{i} = e \, f_{ij} \, \dot{x}^{j} \, , \;\;\;\; (i=1,2,\cdots,D-1) \,,
\label{geod2}
\end{equation}
i.e. the $(D-1)$-dimensional Euclidean Lorentz force equation for a charged
point particle of charge/mass ratio equal to $e$. Moreover, contracting 
(\ref{geod2}) further by $\dot{x}^{i}$, one obtains a second constant of 
motion \( \dot{x}^{i} \, \dot{x}^{i} = \ell^{2} = constant . \)
Since
\( g_{\mu\nu} \, \dot{x}^{\mu} \, \dot{x}^{\nu} =
h_{\mu\nu} \, \dot{x}^{\mu} \, \dot{x}^{\nu} - (u_{\mu} \, \dot{x}^{\mu})^{2}
= \ell^{2} - e^2 , \)
one concludes that the nature of the geodesics necessarily depends on the
sign of $\ell^{2}-e^2$.

In retrospect, we have shown that the G{\"o}del-type metric (\ref{met})
solves the Einstein-Maxwell dust field equations in $D$ dimensions
provided the flat $(D-1)$-dimensional Euclidean source-free Maxwell's
equation (\ref{max}) holds. Moreover the geodesics of the G{\"o}del-type
metrics are described by the $(D-1)$-dimensional Euclidean Lorentz
force equation (\ref{geod2}).

\subsection{\label{spesol} A special solution to (\ref{max})}

A solution to (\ref{max}) is given by the simple choice
\( u_{i}= \frac{b}{2} \, J_{ij} \, x^{j} , \)
where $b$ is a real constant (we keep the 1/2 factor for later convenience)
and $J_{ij}$ is fully antisymmetric with constant components that satisfy
\[J^{k}\,_{j} \, J^{i}\,_{k} = - \delta^{i}_{j} \, . \]
[Of course this is only possible when $D$ is odd.] In this case,
$f_{ij}=b J_{ij}$ and $f_{0 \mu}=0$ as before. Then
\( f_{\mu}\,^{\alpha} \, f_{\nu \alpha} = b^2 \, \bar{\delta}_{\mu\nu}
= b^2 \, h_{\mu\nu} \)
and $f^2=b^2 (D-1)$. Using (\ref{ricci}) and (\ref{riccis})
with $j_{\mu}=0$, the Einstein tensor can be written as
\[ G_{\mu \nu} = \frac{1}{8} \, b^2 \, (5-D) \, g_{\mu \nu}
+ \frac{1}{4} \, b^2 \, (D+1) \, u_{\mu} u_{\nu} \, .\]
This in turn can be interpreted as coming from a perfect fluid source
\[ G_{\mu \nu} = T_{\mu \nu} = p \, g_{\mu \nu}
+ (p+\rho) \, u_{\mu} u_{\nu} \]
by identifying the pressure $p$ of the fluid as
\( p = \frac{1}{8} \, b^2 \, (5-D) \)
and the mass-energy density $\rho$ with
\( \rho = \frac{3}{8} \, b^2 \, (D-1) \, .\)
Notice that in this picture $p=0$ when $D=5$ and $p<0$ when $D>5$.

Alternatively, one can repeat this analysis by writing
\[ h_{\mu \nu} = \frac{1}{b^2} \, f_{\mu}\,^{\alpha} \, f_{\nu \alpha}
\;\;\; \mbox{and} \;\;\;
u_{\mu} u_{\nu} = \frac{1}{b^2} \, f_{\mu}\,^{\alpha} \, f_{\nu \alpha}
- g_{\mu \nu} \]
in (\ref{ricci}). In this case the Einstein tensor
can be written in the form
\[ G_{\mu \nu} = \frac{D+1}{4} \, \left[ f_{\mu}\,^{\alpha}\, f_{\nu \alpha}
- \frac{3}{2(D+1)} \, f^2 \, g_{\mu\nu} \right] \, . \]
If one is to consider this as an Einstein-Maxwell theory so that
$G_{\mu \nu} \sim T^{f}_{\mu \nu}$, then
\[ \frac{3}{2(D+1)} = \frac{1}{4} \]
which yields $D=5$.

As a result, when $D=5$ the special solution given above can either
be thought of as describing a spacetime filled with dust or as a
solution to the Einstein-Maxwell theory. However, in general odd
dimensions it can be considered as a solution of Einstein theory
coupled with a perfect fluid source where the pressure $p<0$ when
$D>5$.

\subsection{\label{withctcs} Spacetimes containing closed timelike curves}

In this subsection we give a simple solution which corresponds to a spacetime
(\ref{met}) that contains closed timelike or null curves. Here we take $D=4$
for simplicity but what follows can easily be generalized to higher dimensions.

Obviously the simple choice $u_{i}=Q_{ij} x^{j}$, where $Q_{ij}$ is fully
antisymmetric with constant components $(i,j=1,2,3)$, solves (\ref{max}).
Now let $Q_{13}=Q_{23}=0$ but $Q_{12} \neq 0$ for simplicity. Then
\[ u_{\mu} dx^{\mu} = dt + Q_{12} (x^2 dx^1 - x^1 dx^2) \, , \]
and employing the ordinary cylindrical coordinates $(\rho,\phi,z)$ this
can be written as
\[ u_{\mu} dx^{\mu} = dt - Q_{12} \, \rho^{2} \, d \phi \, . \]
Using (\ref{met}), this in turn implies that the line element is
\[ ds^2 = d\rho^{2} + \rho^{2} d\phi^{2} + dz^2
- (dt - Q_{12} \, \rho^{2} \, d \phi)^{2} \, . \]

Consider the curve $C=\{(t,\rho,\phi,z) \, | \, t=t_0 \, , \rho=\rho_0
\, , z=z_0 \}$, where $t_0, \rho_0$ and $z_0$ are constants, in the
manifold $M$. The norm of the tangent vector
$v^{\mu}= (\partial/\partial \phi)^{\mu}$
to this curve is then
\[ v^{2} \equiv v_{\mu} v^{\mu} = g_{\phi\phi} =
\rho_{0}^2 \, (1 - (Q_{12})^2 \, \rho_{0}^2) \, . \]

For a spacelike tangent vector, one has $v^2 > 0$, of course. The spacetime
we are studying is obviously homogeneous and there passes a curve like $C$
from each point of such a spacetime. Since $\phi$ is a periodic variable
with $\phi=0$ and $\phi=2 \pi$ identified, one then clearly finds that there
exist closed timelike and null curves for $\rho_{0} \ge 1/|Q_{12}|$ in this
spacetime since then $v^{2} \le 0$. One can also show that there exits no
closed timelike or null geodesics in this geometry.

\subsection{\label{noctcs} Spacetimes without any closed timelike curves}

In this subsection we present a solution which describes a spacetime
(\ref{met}) that doesn't contain any closed timelike or null curves.

Now let $u_{i}=s(x^{j}) \, \omega_{i}$ where 
$\omega^{i}=\delta^{ij} \omega_{j}$ is a constant vector
and $s$ is a smooth function of the spatial coordinates $x^{j}$,
$(i,j=1,2,\cdots,D-1)$. Hence $f_{ij}=(\partial_{i} s) \, \omega_{j}
- (\partial_{j} s) \, \omega_{i}$ and (\ref{max}) gives
\[ \partial_{i} f_{ij} = (\nabla^{2} s) \, \omega_{j}
- (\partial_{i} \partial_{j} s) \, \omega_{i} = 0 \, . \]
Then (\ref{max}) is satisfied if one chooses $\nabla^{2} s=0$ and
$\omega_{i} \, \partial_{i} s= constant$, which can further be set
equal to zero.

Now let us take $D=4$ specifically, but the following discussion
can of course be generalized to $D>4$ as well. As a simple example,
choose $\omega_{i} = \delta_{i}^{3}$ above. Then any function
$s$ which is harmonic in the $(x^{1},x^{2})$ variables will do. Now
\[ u_{\mu} dx^{\mu} = dt + s(x^{1},x^{2}) \, dx^{3} \, , \]
and using the cylindrical coordinates again, the line element becomes
\[ ds^2 = d\rho^{2} + \rho^{2} d\phi^{2} + dz^2
- (dt + s(\rho,\phi) \, dz)^2 \, . \]

Consider the curve $C$ and its tangent vector $v^{\mu}$ we used in Subsection
\ref{withctcs} again. Now the norm of $v^{\mu}$ is
\[ v^{2} = g_{\phi\phi} = \rho_{0}^2 \, , \]
and this is obviously always positive definite, $v^{2} > 0$, i.e.
$v^{\mu}$ is always spacelike. Hence we see that the closed curve we used 
in the previous subsection is no longer timelike.

It is worth pointing out that the consideration of a curve of the form
$\bar{C}=\{(t,\rho,\phi,z) \, | \, t=t_0 \, , \rho=\rho_0
\, , \phi=\phi_0 \, , z \in [0, 2 \pi) \}$, where $t_0, \rho_0$
and $\phi_0$ are constants, and its tangent vector
$\bar{v}^{\mu}=(\partial/\partial z)^{\mu}$ gives
\[ \bar{v}^2 = g_{zz} = 1-(s(\rho_0,\phi_0))^2 \, , \]
which at first sight indicates the existence of closed timelike or null
curves in this geometry. On the other hand we don't confine ourselves
to the small patch of spacetime where the $z$ coordinate is on $S^{1}$,
we are interested in the universal covering of this patch and thus take
$z$ to be on the real line $\mathbb{R}$.

We thus conclude that the solutions we present correspond to spacetimes that
contain both closed timelike and null curves and that contain neither of these
depending on how one solves (\ref{max}).

\section{\label{vardims} Solutions of Various Supergravity Theories 
With Flat Backgrounds}

In this section, we use the results we have obtained so far in
constructing solutions to some supergravity theories in dimensions
$D \ge 5$ with flat backgrounds.

\subsection{\label{5dim} Five dimensions}

The Bosonic part of the minimal supergravity in $D=5$ has the following
field equations \cite{gghp}, \cite{her1}
\begin{eqnarray}
R_{\mu \nu} & = & 2 \, (F_{\mu \alpha}\, F_{\nu}\,^{\alpha}
- \frac{1}{6}\, g_{\mu \nu}\, F^{2}) \;\; \Leftrightarrow \;\;
G_{\mu \nu} = 2 \, T^{F}_{\mu \nu} \, , \label{d=5eq1}\\
\nabla_{\mu}\,F^{\mu \nu} & = & \frac{1}{2\, \sqrt{3}}\,
\eta^{\alpha \beta \gamma \mu \nu}\, F_{\alpha \beta}\, F_{\gamma \mu} \, ,
\label{d=5eq2}
\end{eqnarray}
where the Levi-Civita tensor $\eta$ is given in terms of the Levi-Civita
tensor density $\epsilon$ by \( \eta^{\alpha \beta \gamma \mu \nu} =
\epsilon^{\alpha \beta \gamma \mu \nu} / \sqrt{-g} \).

Let $A_{\mu}= b\, u_{\mu}$, where $b$ is a real constant. One then has
\( F^{\mu \nu} = b \, f^{\mu \nu} .\) Now let us concentrate 
on (\ref{d=5eq2}) first. By (\ref{cmax}),
\( \nabla_{\mu} \, F^{\mu\nu} = \frac{b}{2}\, f^{2}\, u^{\nu}\) and since
\( u^{\nu} = - \delta^{\nu}_{0}\), one finds that
\[ \nabla_{\mu} \, F^{\mu\nu} = - \frac{b}{2}\, f^{2}\, \delta^{\nu}_{0} \, .\]
Now when $\nu = i$, one of the first four indices of $\eta$ on the right hand
side of (\ref{d=5eq2}) has to be 0 and since $F_{0 k}= 0$, (\ref{d=5eq2}) is
satisfied identically in this case. When $\nu = 0$,
\( \nabla_{\mu} \, F^{\mu 0} = - \frac{b}{2}\, f^{2} \). On the other hand,
the right hand side of (\ref{d=5eq2}) gives
\[ \frac{1}{2\, \sqrt{3}}\,
\epsilon^{\alpha \beta \gamma \mu 0}\, F_{\alpha \beta}\, F_{\gamma \mu} =
\frac{b^2}{2\, \sqrt{3}}\, \epsilon^{ijkl}\, f_{ij}\, f_{kl}\]
since $\sqrt{-g}=1$. So if one chooses $f_{ij}$ to be further (anti)
self-dual in Euclidean ${\mathbb R}^4$, i.e.
\( f_{ij} = \pm \frac{1}{2} \epsilon_{ijkl} \, f_{kl}\) (in addition to the
condition (\ref{max})), then the right hand side of (\ref{d=5eq2}) gives
$\pm \frac{b^2}{\sqrt{3}} f^2$ and comparing with the left hand side, one
finds $b=\mp \frac{\sqrt{3}}{2}$.

Paraphrasing, (\ref{d=5eq2}) is satisfied provided $b=\mp \frac{\sqrt{3}}{2}$,
$f_{\mu \nu}$ satisfies (\ref{cmax}) and $f_{ij}$ is (anti) self-dual in the
Euclidean ${\mathbb R}^4$ space. Such an $f_{\mu \nu}$ can easily be
constructed by choosing \(u_{i}=J_{ij}\, x^{j} \;\, (i,j=1,2,3,4)\),
where all components of the fully antisymmetric $J_{ij}$ are constants
with \(J^{k}\,_{j}\, J^{i}\,_{k} = - \delta^{i}_{j}, \)
i.e. $J_{ij}$ defines
an almost complex structure in ${\mathbb R}^4$. Moreover, for $f_{ij}$ to be
(anti) self-dual, $J_{ij}$ must itself be (anti) self-dual in
${\mathbb R}^4$, i.e. \( J_{ij} = \pm \frac{1}{2} \epsilon_{ijkl} \, J_{kl}\).

With the special choice of $u_{i}$ as above,
\[ T^{f}_{\mu \nu} = \frac{1}{4} \, f^2 \, u_{\mu} u_{\nu}
\;\;\; \mbox{and} \;\;\;
T^{F}_{\mu \nu} = b^2 \, T^{f}_{\mu \nu} \, . \]
Looking back at (\ref{d=5eq1}) and using (\ref{eint}) of
Section \ref{godmet}, one finds that
\( T^{f}_{\mu \nu} = \frac{1}{2(4b^2-1)} \, f^2 \, u_{\mu} u_{\nu} \).
Comparing with the $ T^{f}_{\mu \nu}$ above, this again yields
$b=\mp \frac{\sqrt{3}}{2}$ as before. In fact this solution
is supersymmetric \cite{bghv}, \cite{ht}, \cite{gghp}, \cite{her1}.

\subsection{\label{6dim} Six dimensions}

In our conventions, the Bosonic part of the $D=6, N=2$ supergravity theory
\cite{sez} reduces to the following field equations when all the
scalars of the hypermatter $\phi^{a}$ and the 2-form field $B_{\mu \nu}$
in the theory are set to zero and when one assumes the dilaton $\varphi$
to be constant with $\mu \equiv e^{\sqrt{2}\, \varphi}$:
\begin{eqnarray}
R_{\mu \nu} & = & 2 \, \mu \, F_{\mu \rho} \, F_{\nu}\,^{\rho}
+ \mu^2 \, G_{\mu \rho \sigma} \, G_{\nu}\,^{\rho \sigma} \, , \label{d=6eq1}\\
\nabla_{\mu} \, F^{\mu \nu} - \mu \, G^{\nu \rho \sigma}
\, F_{\rho \sigma} & = & 0 \, , \label{d=6eq2}\\
\nabla_{\mu} \, G^{\mu \nu \rho} & = & 0 \, , \label{d=6eq3}\\
\frac{1}{3} \, \mu \, G_{\mu \nu \rho} \, G^{\mu \nu \rho}
+ \frac{1}{2} \, F_{\mu \nu} \, F^{\mu \nu} & = & 0 \, . \label{d=6eq4}
\end{eqnarray}

Here all Greek indices run from 0 to 5 and $G_{\mu \nu \rho}$ is given by
\begin{equation}
G_{\mu \nu \rho} = F_{\mu \nu} \, A_{\rho} + F_{\nu \rho} \, A_{\mu}
+ F_{\rho \mu}\, A_{\nu} \, \label{gdef}
\end{equation}
and instead of a Yang-Mills field, we have taken an ordinary vector
field $A^{\mu}$ to be present. 

Let $A_{\mu} = \lambda u_{\mu}$, where $\lambda$ is a
real constant. Then (\ref{d=6eq3}) is satisfied identically since
\(u^{\mu} = - \delta^{\mu}_{0}\) for our choice. One also finds that
with this $A_{\mu}$,
\( G_{\mu \nu \rho} \, G^{\mu \nu \rho} = -3 \, \lambda^4 \, f^2 \)
and \(F_{\mu \nu} \, F^{\mu \nu} = \lambda^2 \, f^2 . \)
Hence (\ref{d=6eq4}) holds provided $\mu \lambda^2 = 1/2$. Since
\( \nabla_{\mu} \, F^{\mu\nu} = \frac{\lambda}{2}\, f^{2}\, u^{\nu}\) for us
and since
\( G^{\nu \rho \sigma} \, F_{\rho \sigma} = \lambda^3 \, f^{2}\, u^{\nu} ,\)
one again finds that (\ref{d=6eq2}) is satisfied when $\mu \lambda^2 = 1/2$.
Finally, noting that
\[ F_{\mu \rho} \, F_{\nu}\,^{\rho} = \lambda^2 \,
f_{\mu \rho} \, f_{\nu}\,^{\rho} \, , \;\;
G_{\mu \rho \sigma} \, G_{\nu}\,^{\rho \sigma} = \lambda^4 \,
(f^2 \, u_{\mu}\, u_{\nu} - 2 f_{\mu \rho} \, f_{\nu}\,^{\rho}) \]
and using (\ref{ricci}), one finds that (\ref{d=6eq1}) is again satisfied
when $\mu \lambda^2 = 1/2$.

Hence our G{\"o}del-type metric (\ref{met}) and choice of $A_{\mu}$
yield a class of exact solutions to $D=6, N=2$ supergravity
theory. It should be further investigated to see whether this class of
solutions preserves any supersymmetry.

\subsection{\label{8dim} Eight dimensions}

The Bosonic part of the gauged $D=8, N=1$ supergravity theory coupled to
$n$ vector multiplets \cite{sals} has field equations which are very
similar to the field equations of $D=6$, $N=2$ supergravity that we have
examined in Subsection \ref{6dim}. Taking an ordinary vector field instead
of a Yang-Mills field and setting the 2-form field $B_{MN}$ equal to zero
(as was done in Subsection \ref{6dim}), one again has
\begin{equation}
G_{MNP} = F_{MN} \, A_{P} + F_{NP} \, A_{M} + F_{PM}\, A_{N} \, , \label{g8def}
\end{equation}
similarly to (\ref{gdef}), where now capital Latin indices run from 0 to 7. We
also set all the scalars in the theory to zero but assume that the dilaton
$\sigma$ is constant with $\mu \equiv e^{\sigma}$. These assumptions
lead to the following field equations (see (26) of \cite{sals})
\begin{eqnarray}
R_{MN} & = & 2 \, \mu \, F_{MP} \, F_{N}\,^{P}
+ \mu^2 \, G_{MPS} \, G_{N}\,^{PS} \, , \label{d=8eq1}\\
\nabla_{M} \, F^{MN} & = & \mu \, G^{NPS} \, F_{PS} \, , \label{d=8eq2}\\
\nabla_{M} \, G^{MNP} & = & 0 \, , \label{d=8eq3}\\
\frac{2}{9} \, \mu^2 \, G_{MNP} \, G^{MNP}
+ \frac{1}{3} \, \mu \, F_{MN} \, F^{MN} & = & 0 \, . \label{d=8eq4}
\end{eqnarray}
which have the same form as (\ref{d=6eq1}), (\ref{d=6eq2}), (\ref{d=6eq3})
and (\ref{d=6eq4}), respectively.

Letting
\[ g_{MN}= h_{MN} - u_{M} u_{N} \]
(as in Section \ref{godmet}) and $A_{M}= \lambda u_{M}$ (with $\lambda$ real),
and following similar steps as in Subsection \ref{6dim}, it immediately follows
that one obtains exact solutions to gauged $D=8$, $N=1$ supergravity with
matter couplings provided $\mu \lambda^{2}=1/2$. Once again the conditions
on $u_{M}$ under which these solutions are supersymmetric should be
studied further.

\subsection{\label{10dim} Ten dimensions}

The following field equations can be obtained from a five-dimensional action
which is itself obtained by a Kaluza-Klein reduction of the type IIB
supergravity theory with only a dilaton, a Ramond-Ramond 2-form gauge
potential and a graviton. (The details of the reduction process, the
corresponding splitting of the ten-dimensional coordinates and the
metric ansatz employed are explained in detail in \cite{her1} and we
directly make use of the results of that article here.)
\begin{eqnarray}
\nabla_{\mu} \, F^{\mu \nu} & = & - \frac{1}{2} \, H^{\nu \rho \sigma}
\, F_{\rho \sigma} \, , \label{d=10eq1}\\
\nabla_{\mu} \, H^{\mu \nu \rho} & = & 0 \, , \label{d=10eq2}\\
H_{\mu \nu \rho} \, H^{\mu \nu \rho} & = &
- 3 \, F_{\mu \nu} \, F^{\mu \nu} \, , \label{d=10eq3}\\
G_{\mu \nu} & = & F_{\mu \alpha} \, F_{\nu}\,^{\alpha}
- \frac{1}{4} \, g_{\mu \nu} \, F^{\alpha \beta} \, F_{\alpha \beta}
 + \frac{1}{4} \, \left( H_{\mu \alpha \beta} \, H_{\nu}\,^{\alpha \beta}
- \frac{1}{6} \, g_{\mu \nu} \, H_{\rho \alpha \beta} \, H^{\rho \alpha \beta}
\right) \, . \label{d=10eq4}
\end{eqnarray}
Here all Greek indices run from 0 to 4.

Notice the striking resemblance of these equations to the equations of
the $D=6,N=2$ supergravity theory of Subsection \ref{6dim}. We want to see
whether our G{\"o}del-type metric ansatz (\ref{met}) and choice
$A_{\mu} = \lambda u_{\mu}$, with $\lambda$ a constant, solves
equations (\ref{d=10eq1}), (\ref{d=10eq2}), (\ref{d=10eq3}), (\ref{d=10eq4}) 
above. We take the 2-form field $B$ to be
zero to that effect and following \cite{her1} find that $H_{\mu \nu \rho}$
is given by ($H = - A \wedge dA$)
\[ H_{\mu \nu \rho} = - \left( F_{\mu \nu} \, A_{\rho}
+ F_{\nu \rho} \, A_{\mu} + F_{\rho \mu}\, A_{\nu} \right) \]
which already resembles the $G_{\mu \nu \rho}$ of Subsection \ref{6dim}.
Following similar steps to what was done in Subsection \ref{6dim}, one can
easily show that our G{\"o}del-type metric ansatz (\ref{met}) and
choice of $A_{\mu}$ solve equations (\ref{d=10eq1}), (\ref{d=10eq2}), 
(\ref{d=10eq3}), (\ref{d=10eq4}) above provided that $\lambda^2 = 1$.

Following the discussion of \cite{her1}, if one further assumes that
$u_{\mu}$ is chosen in such a way that the 3-form field $H=- \star dA$,
where $\star$ denotes Hodge duality with respect to the G{\"o}del-type 
metric (\ref{met}), and that the gauge field $A$ is rescaled
as $A \rightarrow 2 A/ \sqrt{3}$, this five-dimensional solution can be
further uplifted as the solution
\begin{eqnarray}
ds^2 & = & g_{\mu \nu} dx^{\mu} dx^{\nu} + (dy + \frac{2}{\sqrt{3}} \,
A_{\mu} dx^{\mu})^2 + ds^2({\mathbb T}^4) \, , \label{d=10met}\\
\hat{H} & = & \frac{2}{\sqrt{3}} \, dA \wedge (dy + \frac{2}{\sqrt{3}} A)
- \frac{2}{\sqrt{3}} \star \, dA \, , \label{d=10H}
\end{eqnarray}
of the type IIB supergravity theory. Here $ds^2({\mathbb T}^4)$ is the
metric on a flat four-torus and $y$ denotes one of the singled out extra
dimensions. (See \cite{her1} for details.)

\subsection{\label{11dim} Eleven dimensions}

The solution we gave in Subsection \ref{5dim} can also be uplifted to eleven
dimensions as well \cite{gghp}. The field equations for the Bosonic
part of $D=11$ supergravity are as follows \cite{cjs}:
\begin{eqnarray}
R_{AB} & = & \frac{1}{12} \left( H_{ACDE} \, H_{B}\,^{CDE}
- \frac{1}{12} \, H^2 \, g_{AB}  \right) \, , \label{d=11eq1} \\
\partial_{A} \,(\sqrt{-g} \, H^{ABCD}) & = & \frac{1}{2 \, (4!)^2} \,
\epsilon^{BCDMNKLPRST} \, H_{MNKL} \, H_{PRST} \, . \label{d=11eq2}
\end{eqnarray}
Here capital Latin indices run from 0 to 10. Now split the spacetime into
$x^{A}=(x^{\mu},x^{m})$ where $\mu=0,1,2,3,4$ of Subsection \ref{5dim},
$m=5,6,\cdots,10$ and let $u_{A}=(u_{\mu},0)$. With this choice of $u_{A}$,
take the metric to be of G{\"o}del-type (\ref{met}) with
\begin{equation}
g_{AB}= h_{AB} - u_{A} u_{B} \, . \label{d=11met}
\end{equation}
Next define a 1-form field $A$ as $A_{A}=k u_{A}$ where
$k$ is a real constant. Then $F=dA=k\,f$ where $f$ has components $f_{\mu\nu}$
as in Subsection \ref{5dim}. Moreover one can also define a second 2-form
${\cal F}$ as
\[ {\cal F} = \frac{2}{\sqrt{3}} \, (dx^{5} \wedge dx^{6}
+ dx^{7} \wedge dx^{8} + dx^{9} \wedge dx^{10}) \]
and the 3-form potential ${\cal G}$ as ${\cal G}={\cal F} \wedge A$. Then
the 4-form $H=d {\cal G}=k \, {\cal F} \wedge f$. 
Using the property that \( {\cal F}_{AB} \, f^{BC} = 0\), one then obtains
\[ H_{ACDE} \, H_{B}\,^{CDE} = 3 k^2 \left[ f^2 \, {\cal F}_{AC} \,
{\cal F}_{B}\,^{C} + {\cal F}^2 \, f_{AC} \, f_{B}\,^{C} \right] \, . \]
Notice that the way ${\cal F}$ is constructed implies that
${\cal F}_{AC} \, {\cal F}_{B}\,^{C}= \frac{4}{3} \delta_{AB}$ 
and ${\cal F}^2=8$. Substituting these in (\ref{d=11eq1}), one gets
\[ R^{\mu}\,_{\nu} = 2 k^2 [f^{\mu\sigma} \, f_{\nu\sigma}
- \frac{1}{6} \, f^2 \, \delta_{5}\,^{\mu}\,_{\nu} ] \]
where $\delta_{5}\,^{\mu}\,_{\nu}$ denotes the five-dimensional
Kronecker delta. This is exactly of the same form as (\ref{d=5eq1})
in $D=5$. For the remaining field equation (\ref{d=11eq2}), first note
that $\sqrt{-g}=1$ and
\begin{equation}
\partial_{A} \, H^{ABCD} = k \, ({\cal F}^{BC} \, \partial_{A} \, f^{AD}
+ {\cal F}^{DB} \, \partial_{A} \, f^{AC}
+ {\cal F}^{CD} \, \partial_{A} \, f^{AB}) \label{delH}
\end{equation}
and the way ${\cal F}$ and $f$ are constructed implies that only one of the
terms in the right hand side of (\ref{delH}) survives, say for \( B=\nu \,,
C=2a+1 \,, D= 2a+2 \, (0 \leq \nu \leq 4; \, 2 \leq a \leq 4)\). Then by
(\ref{cmax}) and $u^{\mu} = - \delta^{\mu}_{0}$, (\ref{d=11eq2}) is
equivalent to
\begin{equation}
- \frac{k}{\sqrt{3}} \, f^2 \, \delta^{\nu}_{0} = \frac{1}{2 \, (4!)^2} \,
\epsilon^{\nu \, 2a+1 \, 2a+2 \, MNKLPRST} \, H_{MNKL} \, H_{PRST} \, .
\label{ara1}
\end{equation}
When \(\nu=i \, (1 \leq i \leq 4)\), one of the last eight indices of
$\epsilon$ on the right hand side of (\ref{ara1}) must be a 0 and since
$H_{0MNS}=0$, (\ref{d=11eq2}) is satisfied identically in this case. When
$\nu=0$, the right hand side of (\ref{ara1}) has nonzero contributions
from terms of the form (henceforth $1 \leq i,j,k,l \leq 4$ and
$5 \leq m,n,p,q \leq 10$)
\begin{equation}
\frac{1}{2 \, (4!)^2} \, \epsilon^{0 \, 2a+1 \, 2a+2 \, ijmn \, klpq} 
\, H_{ijmn} \, H_{klpq} \, .
\label{ara2}
\end{equation}
However the way $H$ and ${\cal F}$ are constructed also implies 
that (\ref{ara2}) is equal to
\[ \frac{1}{2 \, (4!)^2} \, \left( \frac{4!}{2! \, 2!} \right)^2 \, 8 \,
\left(\frac{2k}{\sqrt{3}}\right)^2 \, 
\epsilon^{ijkl} \, f_{ij} \, f_{kl} \]
which now must equal to $-\frac{k}{\sqrt{3}} f^2$ from the left hand side
of (\ref{ara1}). Remember that in $D=5$, we chose $f_{ij}$ to be (anti) 
self-dual in Euclidean $\mathbb{R}^4$, which implies that
\[ -\frac{k}{\sqrt{3}} = \pm \frac{2k^2}{3} \]
or $k = \mp \sqrt{3} /2$ for a solution. This is exactly the value of $b$ 
found in $D=5$.

Hence our G{\"o}del-type metric (\ref{d=11met}) and
choice of $A^{A}$ and ${\cal F}$ yield a class of exact solutions
to $D=11$ supergravity theory. [In fact, it has also been shown that this
class preserves 5/8 of the supersymmetry \cite{gghp}.]

\section{\label{nonflat} G{\"o}del-Type Metrics with $(D-1)$-Dimensional
Non-flat Backgrounds}

So far we have assumed that $h_{0\mu}=0$ and $h_{ij}=\bar{\delta}_{ij}$,
the $(D-1)$-dimensional Kronecker delta symbol, in the metric (\ref{met}). 
We have also taken $u_{0}=1$ and $\partial_{0} u_{\alpha}=0$. These 
assumptions simplified the calculation of the Ricci tensor (\ref{ricci}) and
we showed that for the metric (\ref{met}) to be an exact solution 
to the Einstein Maxwell dust field equations in $D$ dimensions, one 
had the $(D-1)$-dimensional Euclidean source-free Maxwell's
equations (\ref{max}) to solve. Now let us take $h_{\mu\nu}$ to be 
a general $(D-1)$-dimensional non-flat spacetime and for simplicity 
take $u_k=1$.

One now finds that $u^{\mu} h_{\mu\nu}=0$ and the inverse of the metric
is given by (\ref{invmet}) again. However the determinant of $g_{\mu\nu}$
is now different, $g=-h$, where $h$ is the determinant of the
$(D-1)\times(D-1)$ submatrix obtained by deleting the $k^{th}$ row and the
$k^{th}$ column of $h_{\mu\nu}$. The new Christoffel symbols of 
$g_{\mu\nu}$ are given by
\begin{equation}
\tilde{\Gamma}^{\mu}\,_{\alpha \beta} = \gamma^{\mu}\,_{\alpha \beta}
+ u^{\mu} \, u_{\sigma} \, \gamma^{\sigma}\,_{\alpha\beta}
+ \frac{1}{2}\, (u_{\alpha} \, f^{\mu}\,_{\beta}
+ u_{\beta} \, f^{\mu}\,_{\alpha}) - \frac{1}{2} \, u^{\mu} \,
(u_{\alpha, \, \beta} + u_{\beta, \, \alpha}) \, ,
\label{tilchris}
\end{equation}
where $\gamma^{\mu}\,_{\alpha \beta}$ are the Christoffel symbols 
of $h_{\mu\nu}$ and we assume that the indices of $u_{\mu}$
and $f_{\alpha\beta}$ are raised and lowered by the metric $g_{\mu\nu}$. 
By using a vertical stroke to denote a covariant derivative with 
respect to $h_{\mu\nu}$ so that
\( u_{\alpha \vert \beta} = u_{\alpha, \, \beta}
- \gamma^{\nu}\,_{\alpha\beta} \, u_{\nu} ,  \)
(\ref{tilchris}) can be simply written as
\begin{equation}
\tilde{\Gamma}^{\mu}\,_{\alpha \beta} = \gamma^{\mu}\,_{\alpha \beta}
+ \frac{1}{2}\, (u_{\alpha} \, f^{\mu}\,_{\beta}
+ u_{\beta} \, f^{\mu}\,_{\alpha}) - \frac{1}{2} \, u^{\mu} \,
(u_{\alpha \vert \beta} + u_{\beta \vert \alpha}) \, . \label{tilchr}
\end{equation}
Thus the ordinary commas in (\ref{chris}) have been replaced with vertical
strokes and the Christoffel symbols of $h_{\mu\nu}$ have
been added to obtain the Christoffel symbols (\ref{tilchr}) of $g_{\mu\nu}$.

To further remove any ambiguity, let us also denote a covariant derivative
with respect to $g_{\mu\nu}$ by $\tilde{\nabla}_{\mu}$, thus
\[ \tilde{\nabla}_{\beta} \, u_{\alpha} = u_{\alpha, \, \beta}
- \tilde{\Gamma}^{\nu}\,_{\beta\alpha} \, u_{\nu} \, . \]
Using these preliminaries one can in fact show that
\( u^{\alpha} \, \tilde{\nabla}_{\alpha} \, u_{\beta} = 0 \) and 
\( \tilde{\nabla}_{\alpha} \, u_{\beta} = \frac{1}{2} f_{\alpha\beta}, \)
hence $u^{\mu}$ is still tangent to a timelike geodesic curve and is still 
a timelike Killing vector.

The Ricci tensor turns out to be
\begin{equation}
\tilde{R}_{\mu \nu} = \hat{r}_{\mu\nu} + \frac{1}{2} \, f_{\mu}\,^{\alpha} \,
f_{\nu \alpha} + \frac{1}{2}\,(u_{\mu}\, \tilde{j}_{\nu} +
u_{\nu}\, \tilde{j}_{\mu}) +\frac{1}{4}\,f^2 \, u_{\mu} \, u_{\nu} \, ,
\label{tilric}
\end{equation}
where $f^2 = f^{\alpha \beta}\,f_{\alpha \beta}$ as before,
\( \tilde{j}_{\mu} \equiv f^{\alpha}\,_{\mu \vert \alpha} \) and
$\hat{r}_{\mu\nu}$ is the Ricci tensor of $h_{\mu\nu}$.
The Ricci scalar is now readily obtained as
\[ \tilde{R} = \hat{r} + \frac{1}{4}\, f^2 + u^{\mu} \, \tilde{j}_{\mu} \; , \]
where $\hat{r}$ denotes the Ricci scalar of $h_{\mu\nu}$.
[Note that
\( \hat{r} = g^{\alpha\beta} \, \hat{r}_{\alpha\beta} =
\bar{h}^{\alpha\beta} \, \hat{r}_{\alpha\beta} \)
by using $u^{\mu}=-\delta^{\mu}_{k}$, (\ref{invmet}) and 
\( u^{\mu} \gamma^{\nu}\,_{\mu\alpha} = 0 \)
in the explicit calculation of $\hat{r}$.]
Setting $\tilde{j}_{\mu}=0$, the Einstein tensor is found to be
\begin{equation}
\tilde{G}_{\mu\nu} = 
\hat{r}_{\mu\nu} - \frac{1}{2} \, h_{\mu\nu} \, \hat{r}
+ \frac{1}{2}\, T^{f}_{\mu \nu}
+ \left ( \frac{1}{4}\, f^{2} + \frac{1}{2} \, \hat{r} \right)
u_{\mu} \, u_{\nu} \, , \label{tilein}
\end{equation}
where $T^{f}_{\mu \nu}$ denotes the Maxwell energy-momentum tensor for
$f_{\mu\nu}$ as before.

Note that in fact
\[ \tilde{j}_{\mu} = (g^{\alpha\beta} \, f_{\beta\mu})_{\vert \alpha}
= (\bar{h}^{\alpha\beta} \, f_{\beta\mu})_{\vert \alpha} \, . \]
This follows by using (\ref{invmet}), 
\( u^{\alpha} f_{\mu\alpha} = 0 \), $u^{\mu}=-\delta^{\mu}_{k}$
and the initial assumptions on $h_{\mu\nu}$. Hence $\tilde{j}_{\mu}=0$
equivalently implies that $\bar{h}^{\mu\nu} \tilde{j}_{\nu} =0$ or
\begin{equation}
\partial_{\alpha} (\bar{h}^{\alpha\mu} \, \bar{h}^{\beta\nu} \, \sqrt{|h|} \,
f_{\mu\nu}) = 0 \, . \label{feqn}
\end{equation}

Hence we find that the Einstein tensor of the $(D-1)$-dimensional
background $h_{\mu\nu}$ acts as a source term for the Einstein
equations obtained for the $D$-dimensional G{\"o}del-type
metric and that the curvature scalar of the background contributes
to the energy density of the dust fluid provided that the
$(D-1)$-dimensional source-free Maxwell equation (\ref{feqn}) in
the background holds. In the next subsection we give a class
of such solutions in the background of some spaces of constant
curvature.

Note that all the theories we discussed in Subsections \ref{6dim} to
\ref{10dim} have G{\"o}del-type metrics as exact solutions with 
the Ricci flat background metric $h_{\mu \nu}$ where the three form 
field $H_{\mu \nu \alpha}$ and the two form field $F_{\mu \nu}$ are 
given in exactly the same way as those defined in these subsections, 
the dilaton field is taken to be zero and the vector field $u_{\alpha}$ 
now satisfies the Maxwell equation (\ref{feqn}) in the background 
$h_{\mu \nu}$. Hence the Bosonic field equations of all of these 
supergravity theories have effectively reduced to the Maxwell equation
(\ref{feqn})! In Subsection \ref{ssd=3}, we will present solutions of
this type by taking the $(D-1)$-dimensional Tangherlini solution as
the background $h_{\mu \nu}$.

\subsection{\label{conflat} Solutions with $(D-1)$-dimensional
conformally flat backgrounds}

Let us now take the special fixed coordinate $x^{k}$ as $x^{0}$ (i.e. $k=0$)
and let the background $h_{ij}$ be conformally flat so that
$h_{ij}=e^{2 \psi} \bar{\delta}_{ij}$. Here Latin indices run from 1 to $D-1$.
If we denote the radial distance of $\mathbb{R}^{D-1}$ by
$r=\sqrt{x^{i} x^{i}}$, take $\psi=\psi(r)$ and use a prime to denote the
derivative with respect to $r$, then one finds that (see e.g. \cite{wald})
\[ \hat{r}_{ij} - \frac{1}{2} \, h_{ij} \, \hat{r} = \bar{\delta}_{ij}
\left( \psi^{\prime\prime} + \frac{\psi^{\prime}}{r} \right) + x_{i} x_{j}
\left( (\frac{\psi^{\prime}}{r})^2 - \frac{\psi^{\prime\prime}}{r^2} +
\frac{\psi^{\prime}}{r^3} \right) \, , \]
and
\[ \hat{r} = -2 \, e^{-2 \psi} \left( 2 \nabla^2 \psi + (\nabla \psi)^2 \right)
= -2 \, e^{-2 \psi} \left( \frac{2}{r^2} (r^2 \psi^{\prime})^{\prime}
+ (\psi^{\prime})^2 \right) \, , \]
for the special choice $D=4$. [The discussion we present here can easily be
generalized to $D>4$ as well.]

If $\psi^{\prime\prime} + \frac{\psi^{\prime}}{r} = 0$, then one finds that
$\psi = a \ln{r} + b$ for some constants $a$ and $b$. Taking $b=0$, this gives
\[ \hat{r}_{ij} - \frac{1}{2} \, h_{ij} \, \hat{r} =
\frac{a(a+2)}{r^4} x_{i} x_{j} \, . \]
If one chooses $a=-2$, then $h_{ij}= \frac{1}{r^4} \bar{\delta}_{ij}$ and
now both $\hat{r}_{ij}$ and $\hat{r}$ vanish. One now has to solve
(\ref{feqn}) in this background to find the G{\"o}del-type metric
$g_{\mu\nu}$ which solves (\ref{tilein}) with $\hat{r}_{\mu\nu}=\hat{r}=0$.
To construct such a solution, take $u_{i}=s(r) Q_{ij} x^{j}$ where
$Q_{ij}$ is fully antisymmetric with constant components. (\ref{feqn}) 
implies that
\( (r^2 s^{\prime\prime} + 6 r s^{\prime} + 4 s) Q_{j\ell} x^{\ell} = 0 , \)
and in general one obtains
\[ u_{i} = \left( \frac{A}{r^4} + \frac{B}{r} \right) Q_{i\ell} x^{\ell} \]
for some real constants $A$ and $B$.

Thus the line element corresponding to the G{\"o}del-type metric in $D=4$
in this three-dimensional conformally flat background
\[ ds^2 = \frac{1}{r^4} (dx^2 + dy^2 + dz^2) - \left( dt + \left(
\frac{A}{r^4} + \frac{B}{r} \right) Q_{i\ell} x^{\ell} dx^{i} \right)^2 \]
solves the $D=4$ Einstein charged dust field equations.

Let us further set $Q_{13}=Q_{23}=0$ but $Q_{12} \neq 0$ for simplicity
and write the resultant line element using cylindrical coordinates.
One finds
\[ ds^2 = \frac{1}{r^4} (d\rho^2 + \rho^2 d\phi^2 + dz^2) - \left( dt -
Q_{12} \, \frac{\rho^2}{r^4} \, (A + B r^3) \, d\phi \right)^2 \, . \]
Employing the curve $C$ of Subsection \ref{withctcs} and its tangent vector
$v^{\mu}$, one finds that
\[ v^2 = g_{\phi\phi} = \frac{\rho_{0}^2}{r_{0}^4} \left( 1 -
(Q_{12})^2 \, \frac{\rho_{0}^2}{r_{0}^4} \, (A + B r_{0}^3)^2 \right) \]
with $r_{0}^2=\rho_{0}^2 + z_{0}^2$. Since $v^{2}$ is not positive definite
in its full generality, we conclude that there exist closed timelike and
closed null curves in this spacetime.

\subsection{\label{ssiki} Solutions with $(D-1)$-dimensional Einstein spaces
as backgrounds (which are themselves conformally flat)}

Let us again take the special fixed coordinate $x^{k}$ as $x^{0}$ and the
background $h_{ij}$ to be conformally flat so that
$h_{ij}=e^{2 \psi} \bar{\delta}_{ij}$. However let us now assume that
$\hat{r}_{ij} = \Lambda h_{ij}$, where $\Lambda$ denotes the
cosmological constant; i.e. the background
is a $(D-1)$-dimensional Einstein space as well. This yields
\[ \hat{r} = (D-1) \Lambda \quad \mbox{and} \quad
\hat{r}_{ij} - \frac{1}{2} \, h_{ij} \, \hat{r} =
\left( \frac{3-D}{2} \right) \, \Lambda \,
e^{2 \psi} \, \bar{\delta}_{ij} \, . \]
Substituting these in (\ref{tilein}), one finds that when the
background $h_{\mu\nu}$ is a $(D-1)$-dimensional Einstein space, the
G{\"o}del-type metric $g_{\mu\nu}$ provides a solution to
\[ \tilde{G}_{\mu\nu} = \left( \frac{3-D}{2} \right)
\, \Lambda \, g_{\mu\nu} + \frac{1}{2}\, T^{f}_{\mu \nu}
+ \left( \frac{1}{4}\, f^{2} + \Lambda \right) u_{\mu} \, u_{\nu} \, , \]
which describes a charged perfect fluid source with
\[ p = \frac{1}{2} \, (3-D) \, \Lambda \;\;\;\; \mbox{and}
\;\;\;\; \rho = \frac{1}{4}\, f^{2} + \frac{1}{2} \, (D-1) \, \Lambda \, .\]
[For $D \ge 4$, $\Lambda$ must be negative in order to have a positive
pressure density $p$.]

We now further assume that $\psi=\psi(z)$ and take $D=4$ for simplicity.
[One can again generalize the arguments we present here to $D>4$.] Such
a choice of $\psi$ yields
\[ \hat{r}_{ij} - \frac{1}{2} \, h_{ij} \, \hat{r} =
- \partial_{i} \partial_{j} \psi + (\partial_{i} \psi) (\partial_{j} \psi) +
\bar{\delta}_{ij} \nabla^{2} \psi = (-\psi^{\prime\prime} + (\psi^{\prime})^2)
\, \bar{\delta}_{i}^{3} \, \bar{\delta}_{j}^{3} + \psi^{\prime\prime} \,
\bar{\delta}_{ij} \, , \]
where a prime denotes the derivative with respect to $z$.
One then has to solve $-\psi^{\prime\prime} + (\psi^{\prime})^2 = 0$ which
yields $\psi=b-\ln{|z+a|}$ for some real constants $a$ and $b$. Further
demanding $\hat{r}=3 \Lambda$ fixes the constant $b$ so that
\[ h_{ij} = \frac{-2}{\Lambda (z+a)^2} \, \bar{\delta}_{ij} \, . \]
Thus for a physically acceptable background, it must be that $\Lambda<0$.

Hence one now has to solve (\ref{feqn}) in this background so that
the G{\"o}del-type metric $g_{\mu\nu}$ solves (\ref{tilein}) in the form
\[ \tilde{G}_{\mu\nu} =  - \frac{1}{2} \, \Lambda \, g_{\mu\nu}
+ \frac{1}{2}\, T^{f}_{\mu \nu}
+ \left( \frac{1}{4}\, f^{2} + \Lambda \right) u_{\mu} \, u_{\nu} \, , \]
i.e. the charged perfect fluid source has pressure density
\( p = - \frac{1}{2} \, \Lambda , \)
(and $p>0$ when $\Lambda<0$) and energy density
\( \rho = \frac{1}{4}\, f^{2} + \frac{3}{2} \, \Lambda , \)
and $\Lambda<0$ must be chosen properly so that $\rho>0$.

To find a solution to (\ref{feqn}), which simply takes the form
\( \partial_{i} ((z+a) f_{ij}) = 0 \)
in this background, let us use the ansatz
$u_{i}=\bar{\delta}_{i}^{3} s(x,y,z)$. Then
\( f_{ij}=\bar{\delta}_{j}^{3} \, \partial_{i} s
- \bar{\delta}_{i}^{3} \, \partial_{j} s \)
and when the free index $j$ is equal to 3, one finds
\( \partial_{\ell} ((z+a) \, \partial_{\ell} s) = 0 ,\)
where the index $\ell$ runs over 1 and 2. When $j=\ell \neq 3$,
one similarly obtains
\( \partial_{3} ((z+a) \, \partial_{\ell} s) = 0 , \)
which is easily integrated to give
\[ \partial_{\ell} s = \frac{c_{\ell}(x,y)}{z+a} \]
for some ``integration constants'' $c_{\ell}(x,y)$. Consistency with the
$j=3$ equation above further constrains $c_{\ell}$ to satisfy
$\partial_{\ell} c_{\ell}=0$. Hence letting $c_{\ell}=\partial_{\ell} c$
for a potential $c(x,y)$, one finds that
\[ s(x,y,z) = \frac{c(x,y)}{z+a} \]
for a function $c(x,y)$ which is harmonic in the $(x,y)$ variables.

Substituting these into the metric $g_{\mu\nu}$, one finds that the line
element corresponding to this $D=4$ example is given in cylindrical
coordinates as
\[ ds^2 = \frac{-2}{\Lambda (z+a)^2} \, (d\rho^2 + \rho^2 d\phi^2 + dz^2)
- \left( dt + \frac{c(x,y)}{z+a} \, dz \right)^2 \, . \]
One then finds that the norm of the tangent vector $v^{\mu}$ to the curve
$C$ of Subsection \ref{withctcs} is
\[ v^2 = g_{\phi\phi} = \frac{-2 \, \rho_{0}^2}{\Lambda (z_{0}+a)^2} \, . \]
One again sees that for $v^2$ to be positive definite $\Lambda$ must be
$\Lambda<0$. [In that case the pressure density of the perfect fluid is
also positive, $p>0$.] So we see that the closed curve of Subsection 
\ref{withctcs} may be timelike and also the discussion we give at
the end of Subsection \ref{noctcs} regarding the universal covering can
similarly be repeated here.

\subsection{\label{ssd=3} Spacetimes with $(D-1)$-dimensional Riemannian
Tangherlini solutions as backgrounds}

Let the $(D-1)$-dimensional background metric $h_{\mu\nu}$ be the metric of
an Einstein space,
\[ \hat{r}_{\mu\nu}=\frac{2 \Lambda}{3-D} h_{\mu\nu} \,, \;\;\;
(D \neq 3) \,. \]
Then the full
$D$-dimensional Einstein tensor becomes
\[ \tilde{G}_{\mu\nu} = \frac{1}{2}\, T^{f}_{\mu \nu} +
\Lambda \, g_{\mu\nu} + \left( \frac{1}{4}\, f^{2} +
\frac{2 \Lambda}{3-D} \right) u_{\mu} \, u_{\nu} \, , \;\;\;  (D \neq 3) \, .\]
Hence our metric (\ref{met}) solves the Einstein field equations with
a charged dust source and a cosmological constant provided the source-free 
Maxwell equation (\ref{feqn}) holds. The energy density of the dust is
\[ \rho = \frac{1}{4}\, f^{2} + \frac{D-1}{3-D} \, \Lambda \, , \]
where $\Lambda$ must be chosen so that $\rho>0$.

Consider the line element corresponding to the $(D-1)$-dimensional Riemannian
Tangherlini solution with a cosmological constant $\Lambda$
\[ ds_{D-1}^{2} = \zeta dt^2 + \frac{dr^2}{\zeta} + r^2
\, d\Omega_{D-3}^{2} \, , \]
where $\zeta=1-2V$ with
\begin{eqnarray*}
V & = & \left \{\begin{array}{ll} m r^{4-D}
+ \frac{\Lambda}{(D-3)(D-2)}\, r^2,
& (\mbox{$ D \ge 5$})\\
m + \frac{\Lambda}{2}\,r^2,
& (\mbox{$D=4$})
\end{array}
\right. \, ,
\end{eqnarray*}
$m$ is the constant mass parameter, $d\Omega_{D-3}^{2}$ is the metric on the
$(D-3)$-dimensional unit sphere and we take the static limit so that
all acceleration parameters vanish \cite{gs3}.

Let the special fixed coordinate $x^{k}$ be $x^{D-1}$ this time. Let us
also assume that
\( u_{\mu}=u(r) \, \delta_{\mu}^{0} + \delta_{\mu}^{D-1} . \)
Then
\( f_{\mu\nu} = (\delta_{\mu}^{r} \, \delta_{\nu}^{0} -
\delta_{\mu}^{0} \, \delta_{\nu}^{r}) \, u^{\prime} , \)
where a prime denotes the derivative with respect to $r$. The only nontrivial
component of (\ref{feqn}) is obtained when $\beta=0$ and in that case
\( (r^{D-3} \, u^{\prime})^{\prime} = 0 , \)
which yields
\begin{eqnarray*}
u(r) & = & \left \{\begin{array}{ll} a r^{4-D} + b,
& (\mbox{$ D \ge 5$})\\
a \ln{r} + b,
& (\mbox{$D=4$})
\end{array}
\right. \, ,
\end{eqnarray*}
for some real constants $a$ and $b$. Here $b$ is irrelevant since it
can be gauged away and taken as zero.

Substituting these in the metric (\ref{met}), the
$D$-dimensional line element becomes
\[ ds^2 = \zeta dt^2 + \frac{dr^2}{\zeta} + r^2 \, d\Omega_{D-3}^{2}
- (u(r) \, dt + dx^{D-1})^2 \, . \]

For this solution one has
\begin{eqnarray*}
f^2 & = & \left \{\begin{array}{ll} 2\, a^2 \, (D-4)^2 \, r^{6-2D},
& (\mbox{$ D \ge 5$})\\
2 \, a^2 /r^2
& (\mbox{$D=4$})
\end{array}
\right. \, ,
\end{eqnarray*}
and one finds that the energy density of the dust diverges
at $r=0$. In the simple case $a=0$, the Maxwell part of the full energy
momentum tensor vanishes and one just has a dust source with
\[ \rho = \frac{D-1}{3-D} \, \Lambda \, , \]
and for $\rho>0$, $\Lambda$ must be negative. For $D=4$, $\zeta=0$
when $r^2=(1-2m)/\Lambda$.

Notice that the Tangherlini solutions we start with are locally 
Riemannian metrics and the parameter $m$ is no longer the ``mass 
constant'' in the $D$-dimensional spacetime $M$. The local coordinates 
chosen here are $(t, r, \theta_{1}, \theta_{2}, \cdots, \theta_{D-3}, 
x^{D-1})$. Here $x^{D-1}$ plays the role of the ``time'' coordinate 
and the rest of the coordinates $(t, r, \theta_{1}, \theta_{2}, \cdots, 
\theta_{D-3})$ are the $(D-1)$-dimensional cylindrical coordinates. 
Here the $t= constant$ surfaces are planes perpendicular to the $t$-axis 
and the $r=constant$ surfaces are the cylinders containing the set of 
points $r=0$, i.e. the $t$-axis. Hence in our solution the set of points
$\zeta=0$ define a $(D-1)$-dimensional cylinder. As an illustration, 
when $D=5$ and $\Lambda=0$, we have $\zeta=1-\frac{2m}{r}$ and
\[ ds^2 = (1-\frac{2m}{r}) \, dt^2 + \frac{dr^2}{1-\frac{2m}{r}} +
r^2 \, (d\theta^{2} + \sin^{2}{\theta} \, d\phi^2) 
- (\frac{a}{r} \, dt + dx^{4})^2 \, . \]
with $f^2=\frac{2a^2}{r^4}$. Hence at $r=2m$ (a cylinder in 5-dimensions), 
$\zeta=0$. This is not a spacetime singularity and does not describe 
an event horizon either. Inside the cylinder $(r <2m)$, the signature 
of the spacetime changes from $(+,+,+,+,-)$ to $(-,-,+,+,-)$. The 
spacetime singularity is located at $r=0$ (inside the cylinder) which is 
the $t$-axis. It is clear that the interior region of this cylinder is 
not physical. The solutions we give by using the Tangherlini metrics
describe physical spacetimes only in those regions where $\zeta>0$. If 
the $t$-coordinate is assumed to be closed $(t \in (0, 2\pi))$, the 
cylinders mentioned above should be replaced by tori.

As we pointed out earlier, the solutions presented here are also
solutions of the supergravity theories listed in Section \ref{vardims} 
with the two- and three-form fields defined in exactly the same way as
those given in Subsections \ref{6dim} to \ref{10dim} and with a vanishing
dilaton field. The only field equation we had to solve was the Maxwell
equation (\ref{feqn}) in the Riemannian Tangherlini background.

\section{\label{conc} Conclusion}

We have introduced and used G{\"o}del-type metrics to find charged
dust solutions to the Einstein field equations in $D$ dimensions. We started
with a $(D-1)$-dimensional Riemannian background (which could be taken as
either flat or non-flat) and showed that solutions to $D$-dimensional
Einstein-Maxwell theory with a dust source could be obtained provided the
source-free Maxwell's equation is satisfied in the relevant background.
The corresponding geodesics were also found to be described by the
Lorentz force equation for a charged particle in the background geometry.
We gave examples of spacetimes which contained closed timelike and
closed null curves and others that contained neither of these. We used
the G{\"o}del-type metrics to find exact solutions to various kinds
of supergravity theories. By constructing the two form and three
form fields out of the vector field $u_{\mu}$ and by assuming a vanishing
dilaton field, we demonstrated that the Bosonic field equations of these
supergravity theories could effectively be reduced to a simple source-free 
Maxwell's equation (\ref{feqn}) in the relevant background $h_{\mu\nu}$. 

In the case of non-flat backgrounds, we constructed explicit solutions
for $D=4$ when the background was taken to be conformally flat, an Einstein
space and a Riemannian Tangherlini solution. We showed that the G{\"o}del-type 
metrics described a black-hole-like object depending on the 
parameters in the latter case. We also discussed the existence of 
closed timelike or closed null curves for conformally flat and Einstein 
space backgrounds.

It would be worth studying to see how much of the supersymmetry is preserved
in the solutions we have given to various supergravity theories here and
to further seek whether G{\"o}del-type metrics can be employed in
finding new (possibly supersymmetric) solutions to others that we haven't
considered. Throughout this work, we assumed the component of
$u_{\mu}$ along the fixed special coordinate $x^{k}$ to be constant.
Another interesting point to investigate would be to generalize this
assumption to non-constant $u_{k}$. One would then expect to construct
solutions to the Einstein-Maxwell dilaton 3-form field equations then.
Work along these lines is in progress and we expect to report our
results soon.

\section{\label{acknow} Acknowledgment}

We would like to thank the referees for their constructive criticisms.
This work is partially supported by the Scientific and Technical Research
Council of Turkey and by the Turkish Academy of Sciences.

\end{document}